# A pre-emphasis based on the gradient system transfer function reduces steady-state disruptions in bSSFP imaging caused by residual gradients

Hannah Scholten[a], Herbert Köstler[a], Anne Slawig[b]

[a]University Hospital Würzburg, Department of Diagnostic and Interventional Radiology, Würzburg, Germany

[b]Section Medical Physics, University Clinic and Outpatient Clinic for Radiology, Department for Radiation Medicine, University Hospital Halle (Saale), Halle (Saale), Germany

**E-mail addresses:** scholten_h@ukw.de (H. Scholten), koestler_h@ukw.de (H. Köstler), anne.slawig@uk-halle.de (A. Slawig)

**Abstract**

Purpose: To examine whether an advanced gradient pre-emphasis approach based on the gradient system transfer function (GSTF) can mitigate artifacts caused by residual unbalanced gradients in Cartesian balanced steady-state free precession (bSSFP) imaging with non-linear line-ordering.

Theory and Methods: We implemented a gradient pre-emphasis based on the GSTF for bSSFP sequences with linear, centric and quasi-random ordering of the phase-encoding steps. Signal-, noise- and artifact levels were determined in phantom experiments. Furthermore, we simulated the phase accumulating in every TR interval of a Cartesian bSSFP sequence for the three different line-ordering schemes.

Results: The simulations showed that the phase contribution arising from residual unbalanced phase-encoding gradients are the principal cause of steady-state disruptions in our sequence. In the phantom experiments, the GSTF-based gradient pre-emphasis approach reduced the artifact level in bSSFP images with non-linear line-ordering considerably. Compared to the linearly ordered measurement, the relative artifact intensity difference dropped by up to 89 %.

Conclusion: A GSTF-based pre-emphasis approach can successfully mitigate residual unbalanced gradient artifacts in bSSFP imaging with non-linear line-ordering.

**Key words:** balanced steady-state free precession, TrueFISP, gradient impulse response function, gradient system transfer function, pre-emphasis

# 1 Introduction

The dynamic switching of magnetic gradient fields over the course of an imaging sequence in MRI can lead to the induction of eddy currents in and vibrations of the scanner's gradient coils. Both effects deteriorate the spatial and temporal accuracy of the gradient fields. In certain image acquisitions, for example when using non-Cartesian k-space trajectories, temporal gradient errors significantly degrade the image quality. To counteract eddy current effects, most current clinical MR scanners make use of a pre-equalization of the gradient waveforms. The underlying model of the gradient chain consists of a set of exponential filter functions [1, 2, 3]. However, oscillatory field distortions arising from Lorentz force-induced vibrations of the coils cannot be compensated with such a model.

A more sophisticated gradient pre-emphasis can be achieved through the inversion of a linear, time-invariant (LTI)[1] model of the gradient system [4, 5]. If the MR scanner's gradient chain is assumed to represent an LTI system, it can be characterized by the gradient impulse response function (GIRF), or its Fourier-transform, the gradient system transfer function (GSTF) [6, 7]. The GSTF can be determined through simple phantom measurements [6] or with a field camera consisting of multiple NMR probes [7]. Mechanical coil vibrations appear as resonance peaks or dips in the GSTF [8]. The GSTF can thus provide the information necessary for a more detailed prospective correction of gradient distortions than the exponential filter model described previously [5]. Recent publications have demonstrated the application of a GSTF-based gradient pre-emphasis to single-shot echo-planar imaging [4], and spiral imaging [5, 9, 10]. In both cases, the superior pre-emphasis approach minimized k-space trajectory errors.

Another type of MRI sequence, that is sensitive to temporal gradient imperfections, are balanced steady-state free precession (bSSFP) sequences. They are defined by a zero net gradient moment over each TR interval [11, 12, 13]. Balanced SSFP sequences are particularly suitable for fast imaging, as they yield the highest signal per time unit of all currently known MRI sequences [13]. Unfortunately, they are also prone to hardware imperfections [14, 15] and field inhomogeneities [13]. Off-resonances due to static spatial variations of the magnetic field cause a constant dephasing within each TR, which alters the established steady-state. The signal profile as a function of the accrued phase exhibits a periodic behavior with alternating broad passbands and narrow stopbands, resulting in the so-called banding artifacts [16]. They can be mitigated by proper shimming and shortening TR [17]. If the dephasing varies dynamically, on the other hand, the steady-state of the magnetization will be disturbed, causing varying signal intensities during acquisition, which may generate hypo- or hyperintense artifacts [18, 19], or increase aliasing in undersampled data acquisitions [20]. Such dynamic changes in the phase accrual are mainly caused by eddy currents originating from the time-varying magnetic field gradients used for spatial encoding [14]. They only play a minor role in conventional linear Cartesian data acquisition, but their impact becomes obvious when other line-ordering schemes or non-Cartesian k-space sampling are used. For example, artifacts were encountered when accelerating thoracic imaging by using spiral readout gradients [21], cardiac imaging by undersampling and compressed sensing [18, 22], or angiography by centric out Cartesian ordering schemes [19].

---

[1] Abbreviations: LTI, linear time-invariant; GIRF, gradient impulse response function; GSTF, gradient system transfer function; bSSFP, balanced steady-state free precession

To solve this issue, repeating each phase encoding step at the expense of doubling the scan time was proposed [23]. Other attempts aimed at avoiding high fluctuations between subsequent phase encoding gradients and proposed pairing or grouping of k-space lines with similar gradient areas [14, 19, 20, 21]. However, this approach severely limits the flexibility in readout and undersampling designs and may hinder the achievable sampling efficiency. Moreover, employing smoothly varying encoding schemes is not equally effective for every specific set of sequence parameters, and therefore not generalizable. In radial imaging, the use of increasingly small golden angle variants, called tiny golden angles, proved useful to avoid large gradient fluctuations [24]. Another approach, available for 2D acquisitions only, applies an additional low angle dephasing along the slice-encoding direction in order to produce cancelling of negative and positive dephasing within each voxel [14]. Direct annihilation of gradient errors is also possible if they are known in advance, for example by dedicated compensation gradients. This has recently been demonstrated with the help of calibration scans [25] acquired with a dynamic field camera [26]. The need for an additional scan and special hardware, however, inhibits the applicability of this approach in clinical settings.

In this work, we employed a GSTF-based gradient pre-emphasis [4, 5] for bSSFP sequences with different orderings of the phase encoding steps. We show in simulations and experiments that such an approach can largely overcome disruptions of the steady-state in Cartesian k-space acquisitions, independent of the applied line-ordering, and regardless of the step size between subsequent phase encoding steps.

**2 Theory**

*2.1 GSTF-based gradient pre-emphasis*

Under the assumption that the gradient system of the MR scanner represents a linear and time-invariant (LTI) system, it can be completely described by the gradient system transfer function (GSTF). By means of the GSTF, field fluctuations can be derived from the nominal input gradient waveforms of the sequence [7] (c.f. section 2.3). On the other hand, the GSTF can also be used to optimize the input gradient waveform. By means of a pre-emphasis, deviations of the actually produced time course from the nominally prescribed one can be prevented. The pre-emphasized input gradient $G^{\text{pre}}(t)$ can be calculated by multiplying the spectrum of the nominal waveform $\mathcal{G}^{\text{nom}}(\omega) = \mathcal{F}[G^{\text{nom}}(t)]$ with a filter function $H^{\text{pre}}(\omega)$, followed by an inverse Fourier-transform ($\mathcal{F}^{-1}$) [4]:

$$G^{\text{pre}}(t) = \mathcal{F}^{-1}[\mathcal{G}^{\text{nom}}(\omega) \cdot H^{\text{pre}}(\omega)] \tag{1}$$

In order for the played-out gradient on a specific axis to exactly match the nominal one, the pre-emphasis filter would have to be the inverse of the respective self-term of the GSTF. However, the physical limits imposed by the gradient amplifiers usually render this solution inappropriate. Instead, a low-pass target transfer function $H^{\text{tar}}(\omega)$ may be introduced into the pre-emphasis filter as per Eq. (2). [4]

$$H^{\text{pre}}(\omega) = \frac{H^{\text{tar}}(\omega)}{H^1(\omega)} \tag{2}$$

$H^1(\omega)$ denotes the self-term of the GSTF. Our choice for $H^{\text{tar}}(\omega)$ was a two-sided shifted sigmoid function (also known as Fermi function in physics) as defined in Eq. (3).

$$H^{\text{tar}}(\omega) = \frac{1}{\exp\left(\frac{|\omega| - \omega_{\text{FE}}}{\omega_{\text{FW}}}\right) + 1} \tag{3}$$

The parameters $\omega_{\text{FE}}$ ("Fermi edge") and $\omega_{\text{FW}}$ ("Fermi width") control the broadness and steepness of the passband of $H^{\text{tar}}(\omega)$.

*2.2 Field perturbations and phase accumulation in balanced SSFP*

In order for the signal to reach a steady-state in bSSFP sequences, it is imperative that the phase accumulation $\phi$ over each TR is constant [27]. This condition can be violated by magnetic field perturbations caused by eddy currents or coil vibrations, as detailed in the introduction. Consequently, the magnetization acquires an additional phase $\Delta\phi(n)$ that can vary from repetition $n$ to repetition $n+1$, thus disrupting the steady-state. In general, $\Delta\phi(n)$ also depends on the position and can be an arbitrarily complicated function of the spatial coordinates $x$, $y$, and $z$. However, its main contributions are a spatially constant (0[th] order) term $\Delta\phi^0(n)$, resulting from time-varying shifts of the main magnetic field $\Delta B_0(t)$, and a spatially linear (1[st] order) component $\Delta\phi^1(n, x, y, z)$, arising from perturbations of the magnetic gradient fields $\Delta G_x(t), \Delta G_y(t)$, and $\Delta G_z(t)$. These additional field gradients can either occur when switching a gradient on the same axis (self-terms) or on one of the two orthogonal axes (cross-terms). In this work, we focused on the self-term effects, as they are typically much larger than the cross-term effects. We can therefore decompose $\Delta\phi^1(n, x, y, z)$ into three independent components $\Delta\phi^1(n, x), \Delta\phi^1(n, y), \Delta\phi^1(n, z)$. The total additional phase acquired in the n[th] repetition is then approximated by Eq. (4).

$$\Delta\phi(n) = \Delta\phi^0(n) + \Delta\phi^1(n, x) + \Delta\phi^1(n, y) + \Delta\phi^1(n, z) \tag{4}$$

*2.3 GSTF-based predictions of the phase accumulation*

Under the LTI assumption, the field fluctuations $\Delta B_0(t), \Delta G_x(t), \Delta G_y(t)$, and $\Delta G_z(t)$ can be calculated using the GSTF [7]. The actually played out field components are obtained by a convolution of the nominal gradient waveform with the respective component of the GIRF (in the time domain), i.e. the 0[th] order term or the 1[st] order self-terms. Subsequently, these predictions can be used to calculate the phase accrual for any given time point in the sequence, as stated in Eq. (5) and (6).

$$\Delta\phi^0(t) = \gamma \int_0^t \Delta B_0(\tau) d\tau = \gamma \int_0^t \left[\sum_{l \in x,y,z} h_l^0(\tau) * G_l^{\text{nom}}(\tau)\right] d\tau \tag{5}$$

$$\Delta\phi^1(t, l) = \gamma \int_0^t \Delta G_l(\tau) \cdot l \, d\tau = \gamma \int_0^t \left[h_l^1(\tau) * G_l^{\text{nom}}(\tau) - G_l^{\text{nom}}(\tau)\right] \cdot l \, d\tau, \quad l \in x, y, z \tag{6}$$

Here, $\gamma$ denotes the gyromagnetic ratio, $\{G_l^{\text{nom}}\}_{l \in x,y,z}$ is the nominal gradient input on the physical x-, y-, and z-axis, respectively, and $*$ denotes the convolution operator. The 0[th] order terms of the GIRF of the three gradient axes are represented by $\{h_l^0\}_{l \in x,y,z}$. They characterize changes of the main magnetic field as a response to applying a gradient field in the respective direction. The 1[st] order self-terms of the GIRF $\{h_l^1\}_{l \in x,y,z}$ describe the relation between the input and the actually produced gradients on the respective axis. The time point $t = 0$ refers to the start of the pulse sequence. We used Eq. (5) and (6) to assess the expected phase errors in our bSSFP sequence, which are displayed in Figure 1 for a Cartesian k-space acquisition with linear line-ordering. The calculation of the convolutions was realized through multiplications with the respective terms of the GSTF in the frequency domain.

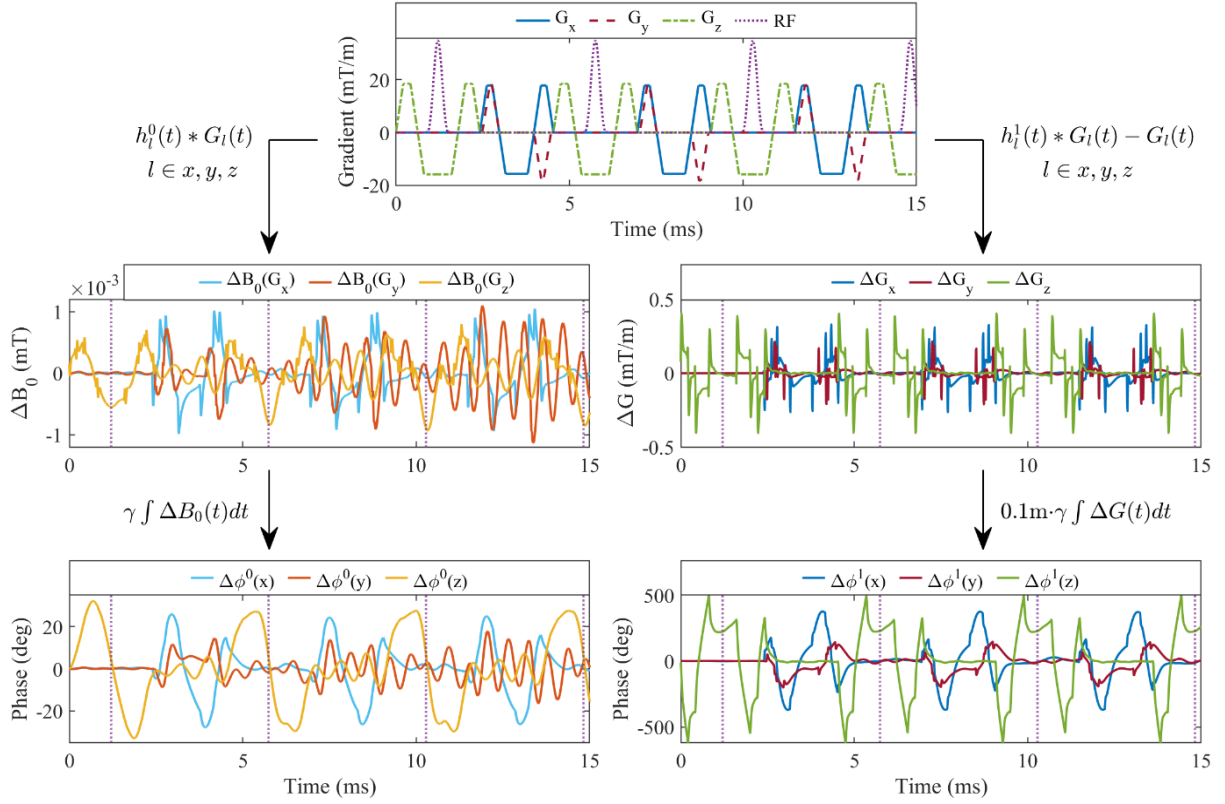

Figure 1. Phase error estimation for our bSSFP sequence for a transverse slice with linear line-ordering. Top: Nominal gradients and RF pulses of the first three TRs. Second row: Field fluctuations in response to the nominal gradients calculated with the $0^{th}$ (left) and $1^{st}$ (right) order terms of the GSTF. Third row: Additional phase accumulations as per Eq. (5) and (6). For the $1^{st}$ order in the right column, a position of 10 cm off-isocenter was assumed in each direction. The dotted vertical lines mark the centers of the RF pulses.

## 3 Materials and Methods

### 3.1 GSTF determination

We used triangular gradient pulses [7] and phantom-based thin-slice measurements [28] to determine the $0^{th}$ order terms and $1^{st}$ order self-terms of the GSTF. Two sets of 16 triangles each and two different measurement schemes were used:

- The first set of triangles had rise times between 10 µs and 160 µs with increments of 10 µs, and a slew rate of 180 mT/m/ms. To capture them, they were played out during the FID readout following a slice-selective excitation [28, 29].
- The second set of triangles had rise times between 280 µs and 430 µs and a slew rate of 150 mT/m/ms (x- and z-axis) or 130 mT/m/ms (y-axis). They were used to detect long-living eddy current effects and field oscillations by shifting the excitation and readout to a timepoint after playing out the triangles [30]. Five measurements of this sort were conducted with the excitation occurring 2 ms, 22 ms, 42 ms, 62 ms, and 82 ms after the start of the triangles. Each readout was 25 ms long, so these measurements had overlapping acquisition windows. A schematic depiction of the sequence diagram can be found in [31].

All six measurement periods together covered a total time window of approximately 107 ms, after discarding spuriously scattered data points at the start and end of each readout (80 µs each). The signal was acquired in two slices of 3 mm thickness, located ± 16.5 mm away from isocenter. The flip angle was set to 90° and TR to 1 s. A spherical phantom of 16.5 cm in diameter and a 16-channel head coil were used. The whole measurement took 6.4 min per gradient axis. The GIRF was calculated in the time domain by solving a linear system of equations, which could be implemented as a matrix equation [32]. A regularization enforcing smoothness at high frequencies in the gradient system's transmission behavior was employed to suppress high frequency noise in the solution, as detailed in [31]. Fourier-transforming the GIRF yielded the GSTF.

*3.2 GSTF-based gradient pre-emphasis*

The pre-emphasized gradient input waveforms were calculated from the nominal gradients according to Eq. (1) and (2), taking into account the 1$^{\text{st}}$ order self-terms $\{H_l^1(\omega)\}_{l \in x,y,z}$. The target transfer function was calculated according to Eq. (3) with $\omega_{\text{FE}} = 2\pi \cdot 15$ kHz and $\omega_{\text{FW}} = 2\pi \cdot 800$ Hz.

Simulations were performed to illustrate the effect of the GSTF-based pre-emphasis on a phase encoding and phase rewinding gradient of opposite polarity. Both gradients had slew rates of 84.4 mT/m/ms and maximum amplitudes of 18.56 mT/m, resulting in a maximum $k$ of 768 000 m$^{-1}$. The rephasing gradient started 0.99 ms after the end of the phase encoding gradient. These parameters correspond to the first phase encoding step in the sequence with linear line-ordering described in section 3.4.

*3.3 Phase accumulation for different line-ordering schemes*

As described in section 2.3, we applied the GSTF to estimate the additional phase contributions that the magnetization would acquire due to hardware imperfections in a bSSFP sequence with the specifications as given in 3.4. We conducted the analysis as depicted in Figure 1 for three different line-ordering schemes of a Cartesian k-space sampling: linear, centric out, and quasi-random. For the latter, a low-discrepancy sequence was employed, as proposed by Niederreiter [33, 34], which holds advantages in gated MR acquisitions because each subset of samples provides high uniformity [35]. This line-ordering is termed "Niederreiter" in the following. For the calculation of the phase accumulation originating from gradient imperfections (i.e. first order terms), as opposed to B$_0$ fluctuations (i.e. zeroth order terms), we assumed a location 10 cm off center in the direction of the respective gradient axis. Subsequently, we considered the change in $\Delta\phi$ from TR to TR (i.e. its derivative with respect to the pulse index $n$) in order to pre-estimate which acquisition would be most prone to artifacts due to steady-state disruptions. A constant additional phase, on the other hand, would not disturb the development of a steady-state, but would influence its nature and therefore the image contrast.

*3.4 bSSFP imaging with GSTF-based gradient pre-emphasis*

To ensure a truly balanced sequence, i.e. the total gradient moment on each axis being zero within each TR interval, we implemented a bSSFP sequence with a GSTF-based gradient pre-emphasis as described in sections 2.1 and 3.2. The pre-emphasized gradient waveforms were calculated on-the-fly during the image acquisition, taking into account effects lasting as long as 3*TR. Consequently, the input for the calculation of the pre-emphasized waveform in repetition $n$ included the gradients in repetitions $n-3$ to $n$. This approach supposedly diminished imperfections in the applied gradients that would result in an additional phase accumulation $\Delta\phi^1(n, x, y, z)$. Imaging experiments were performed with and without the GSTF-based pre-emphasis with all three line-ordering schemes mentioned in section 3.3. We conducted phantom experiments with two cylindrical bottles filled with doped water (diameter: 11.8 cm; volume: 2000 ml; T1 ≈ T2 ≈ 300 ms) that were placed directly on the patient table, oriented in the direction of the scanner bore. Their position was about 5 cm below the magnet's isocenter. We acquired a transversal slice positioned at isocenter with a thickness of 5 mm, a FOV of 300 mm, a matrix size of 256 × 128 with 128 phase-encode steps, a readout bandwidth of 1560 Hz/pixel, TR = 4.54 ms, TE = 2.26 ms, and a flip angle of 38°. Shimming was performed using the scanner's default optimization algorithm for first and second order shim terms. We acquired ten images, the first nine of which served as dummy acquisitions in order to reach the steady-state. Given the chosen slice orientation and phase-encoding direction, the slice selection axis coincided with the physical z-axis of the gradient system, the x-axis corresponded to the frequency encoding direction, and the y-axis to the phase-encoding direction. All measurements were performed on a 3T Siemens MAGNETOM Prisma^fit MRI scanner (Siemens Healthcare, Erlangen, Germany).

In order to quantify the signal-, artifact-, and noise levels in the different images, we defined three distinct regions of interest (ROI) and determined the mean magnitude image intensities within. ROI 1 lies inside the margins of the right bottle. ROI 2 spans a certain artifact-corrupted area outside the two bottles, with a width spanning the whole phantom region in order to capture artifacts from both bottles. ROI 3 is located outside the bottles along the frequency encoding direction. All three ROIs are indicated in Figure 5a.

**4 Results**

*4.1 Measurement of the GSTF*

Figure 2 shows the $0^{th}$ order terms and $1^{st}$ order self-terms of all three physical gradient axes of the GSTF. They were all determined with a frequency resolution of 9.1 Hz, which captures the mechanical resonance peaks in the low frequency range (highlighted in the insets) with a high level of detail. The applied regularization yields virtually noise-free self-terms (c.f. Figure 2b), and moderate noise levels in the $0^{th}$ order terms (c.f. Figure 2a) without the need for averaging.

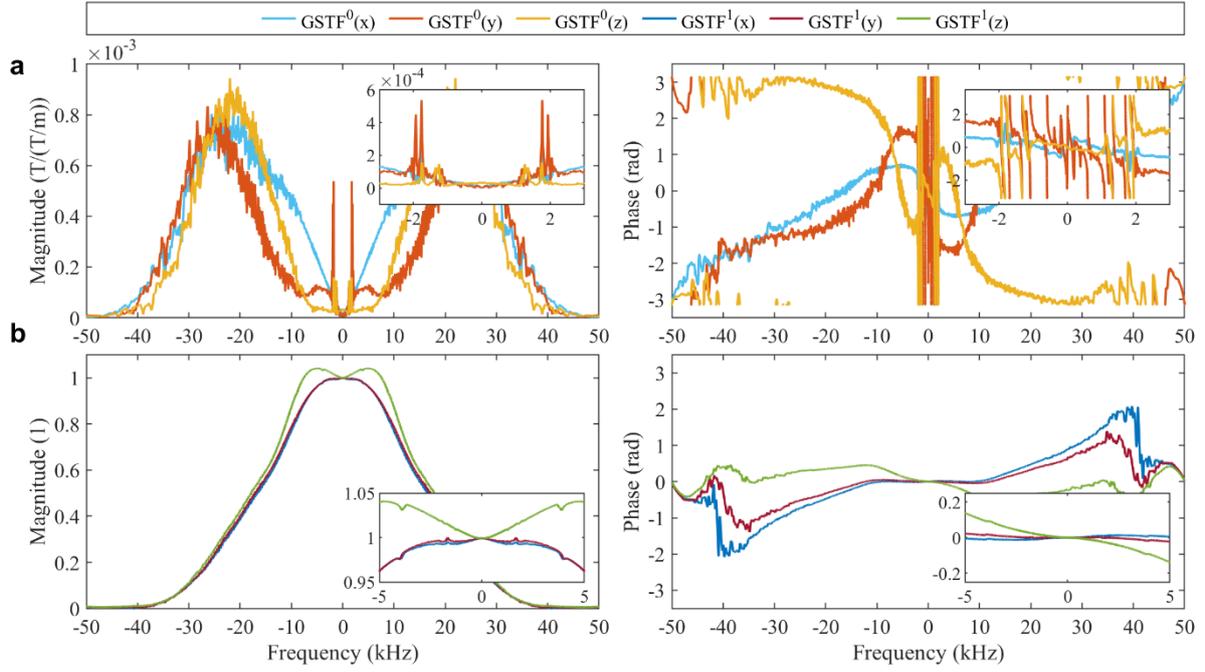

*Figure 2. Magnitude (left) and phase (right) of the gradient system transfer function (GSTF). (a) $0^{th}$ order terms. (b) $1^{st}$ order self-terms. The insets zoom into the low frequency range where mechanical resonance peaks are visible.*

*4.2 Simulated effect of the GSTF-based gradient pre-emphasis*

Figure 3 demonstrates the effect of the GSTF-based pre-emphasis on a pair of phase-encoding gradients. Shown are the temporal gradient evolution and the corresponding k-space coordinate (Figure 3a). In Figure 3b, it can be seen that the gradients exhibit oscillations without the pre-emphasis (dashed blue line), which are counteracted in the prescribed progression with the pre-emphasis (dash-dotted yellow line). The targeted waveform, that is realized by playing out the pre-emphasized one, is shown in Figure 3c (dash-dotted green line). The ripples at the switching points are caused by our choice of the target filter function and the parameters $\omega_{\text{FE}}$ and $\omega_{\text{FW}}$. However, they only live on a very short time scale. Figure 3d displays the k-space coordinates corresponding to the nominal, erroneous, and targeted gradient evolutions. The oscillations present without the GSTF-based pre-emphasis (dashed blue line) are clearly eliminated with it (dash-dotted green line).

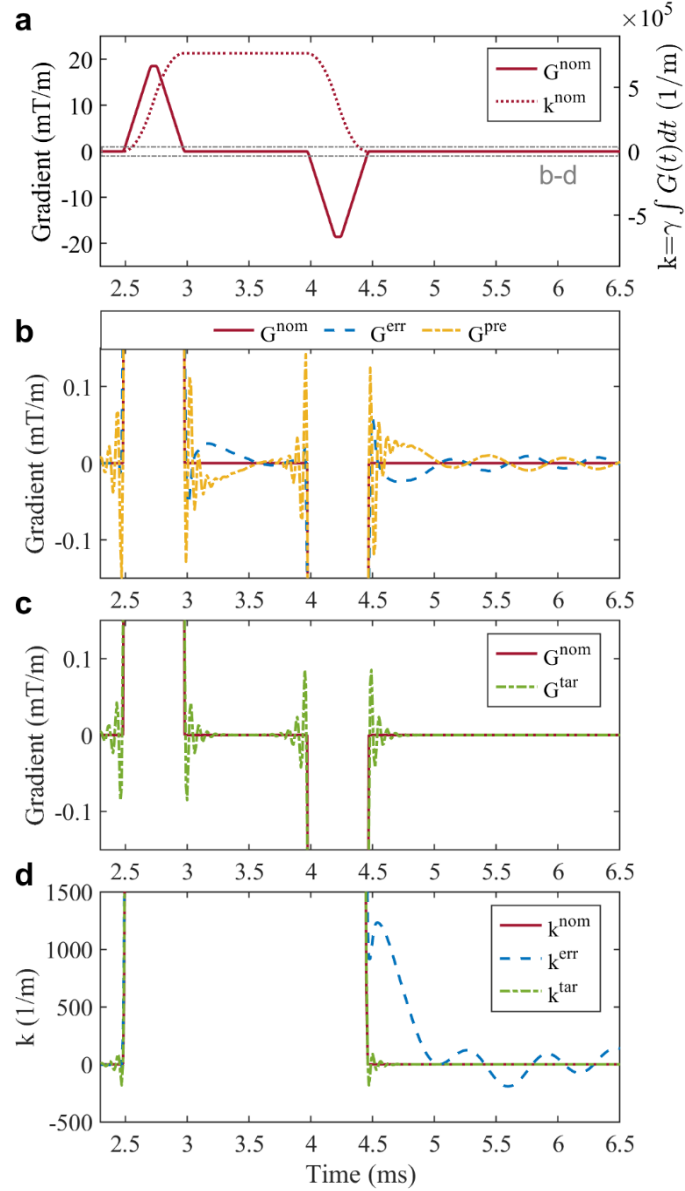

*Figure 3. Time course of a pair of phase encoding gradients (encoder and rewinder) and corresponding k-space coordinate. (a) Nominal gradient and k-space evolution. (b) Nominal, erroneous, and pre-emphasized gradient evolution. The erroneous time course was calculated via multiplication of the nominal one with the GSTF in the frequency domain. The pre-emphasized time course was calculated via Eq. (1). For better visibility of the differences between the curves, the depiction is zoomed in y-direction. (c) Nominal and targeted gradient evolution with the GSTF-based pre-emphasis. The latter was calculated by multiplication of the nominal time course with the target transfer function in the frequency domain. (d) Nominal, erroneous, and targeted evolutions of the k-space coordinate. The depiction is again zoomed in y-direction.*

*4.3 Phase accumulation for different line-ordering schemes*

Figure 4 displays the results of the phase error estimation described in sections 2.3 and 3.3. The phase contributions from the x- and z-axis, i.e. the slice-selection and frequency-encoding direction, are identical for all three line-ordering schemes. They all reach a nearly constant value after a few repetitions (c.f. left column of Figure 4). The smallest contribution in terms of absolute values is given by $\Delta\phi^0(n,x)$, the largest by $\Delta\phi^0(n,z)$. The terms $\Delta\phi^1(n,x)$ and $\Delta\phi^1(n,z)$ yield almost the same additional phase. Concerning the disruptions of the steady state, however, the rate of change from one pulse (n) to the next pulse (n+1) is of more concern than absolute values. As far as the x- and z-axis are concerned, these derivatives of $\Delta\phi^{\{0,1\}}$ with respect to the pulse index $n$, are almost zero over the course of the sequence, once the steady-state is established after the first few pulses in all three line-ordering schemes (c.f. right column of Figure 4). Regarding the phase-encoding direction, both the 0th and 1st order phases, i.e. $\Delta\phi^0(n,y)$ and $\Delta\phi^1(n,y)$, decrease linearly with the pulse number in the linear line-ordering scheme (Figure 4a). For the centric out line-ordering (Figure 4b), the signs of both $\Delta\phi^{\{0,1\}}(n,y)$ and $d(\Delta\phi^{\{0,1\}}(n,y))/dn$ are alternating from pulse to pulse. Their absolute values are increasing with the pulse index. Here, the 1st order phase change $d(\Delta\phi^1(n,y))/dn$ is, on average, about ten times the 0th order phase change $d(\Delta\phi^0(n,y))/dn$ (mean ratio: 10.2 ± 1.6). For the Niederreiter line-ordering (Figure 4c), $\Delta\phi^{\{0,1\}}(n,y)$ and their derivatives show abruptly changing progressions without a clear structure. The absolute value of the ratio of the phase derivatives $d(\Delta\phi^1(n,y))/d(\Delta\phi^0(n,y))$ varies between 1.5 and 421.6 (mean absolute ratio: 29.0 ± 58.9). The change in the 1st order phase contribution thus always exceeds the change in the 0th order contribution.

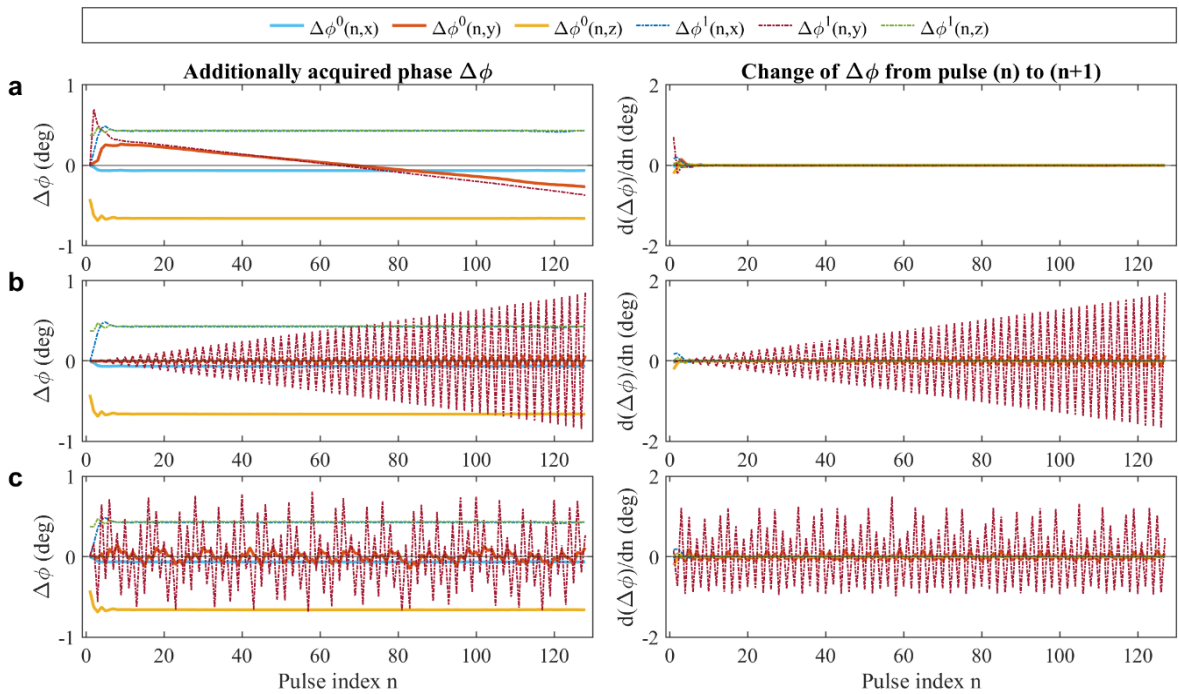

*Figure 4. Phase accrual at the centers of the RF pulses in our bSSFP sequence with different line-ordering schemes. Left: 0th and 1st order components of the additionally acquired phase $\Delta\phi$. Right: Phase change between consecutive RF pulses. (a) Linear line-ordering. (b) Centric out line-ordering. (c) Niederreiter line-ordering.*

*4.4 bSSFP imaging with GSTF-based gradient pre-emphasis*

Figure 5 presents the phantom images acquired with the different line-ordering schemes, with and without the GSTF-based gradient pre-emphasis. With the linear line-ordering (Figure 5a), neither the image without nor the image with pre-emphasis exhibit any visible artifacts. With the centric out (Figure 5b) or Niederreiter line-ordering (Figure 5c) on the other hand, the images without pre-emphasis clearly show artifacts in the phase-encoding (y) direction. With pre-emphasis, the artifacts are greatly reduced for the centric line-ordering. For the Niederreiter line-ordering, they are no longer visible.

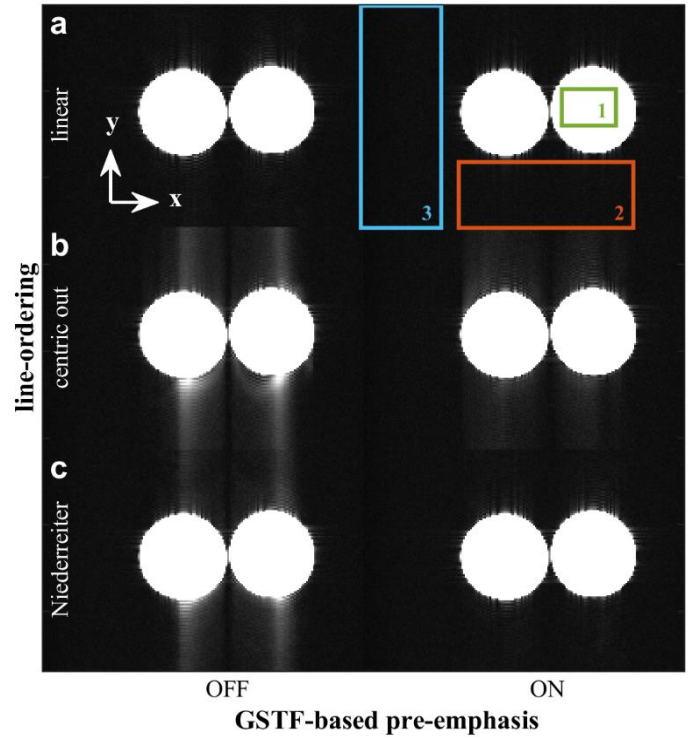

*Figure 5. Transversal bSSFP images of two water bottles, acquired without (left column) and with (right column) the GSTF-based gradient pre-emphasis applied. (a) Linear line-ordering. (b) Centric out line-ordering. (c) Niederreiter line-ordering. The rectangles in (a) mark the regions of interest for the quantification of signal- (ROI 1), artifact- (ROI 2), and noise levels (ROI 3). The arrows indicate the gradient axes.*

A quantitative comparison of intensities in a signal-, an artifact-, and a noise ROI in the images without and with the GSTF-based pre-emphasis for all three ordering schemes of the phase-encoding lines is shown in Figure 6a and b. The values represent the mean magnitude image intensities determined in the ROIs indicated in Figure 5a. In Figure 6a, it can be seen that the signal in the bottles (ROI 1) is very similar in all cases, but increases with the GSTF-based gradient pre-emphasis for the non-linearly ordered acquisition schemes. The noise levels (ROI 3) are the same for all three line-ordering schemes and do not change when the GSTF-based pre-emphasis is applied (Figure 6b). The intensities in the artifact ROI (ROI 2), plotted in Figure 6b, decrease for the non-linear line-orderings when employing the pre-emphasized gradients following the GSTF-based approach. Figure 6c displays the relative signal difference to the linearly ordered measurement with GSTF-based pre-emphasis (right column of Figure 5a) for every respective ROI. This relative signal difference shows virtually no difference for the signal- (ROI 1) and the noise ROI (ROI 3) when comparing the images without and with GSTF-based pre-emphasis. However, in the artifact ROI (ROI 2), the relative intensity difference decreases by 64% for the centric out line-ordering and by 89% for the Niederreiter ordering, when the GSTF-based pre-emphasis is applied.

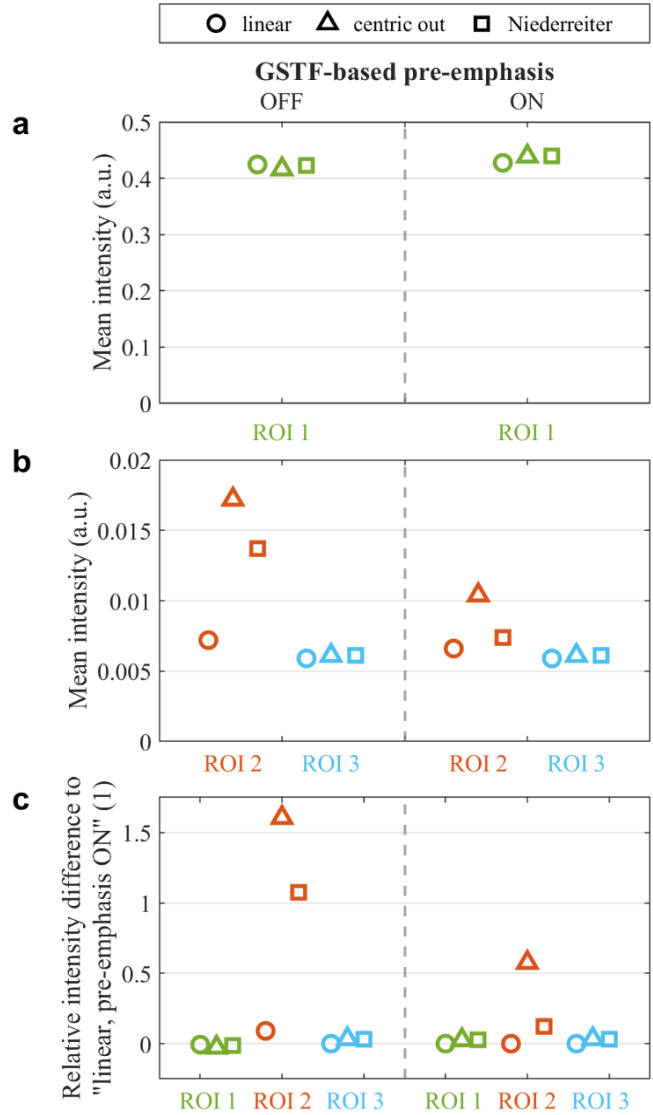

*Figure 6. Quantification of signal-, artifact- and noise levels in the images shown in Figure 5. (a) Mean intensities in ROI 1. (b) Mean intensities in ROIs 2 and 3. (c) ROI-specific relative mean intensity difference to the linearly ordered measurement with GSTF-based pre-emphasis.*

## 5 Discussion

In this work, we explored the impact of a GSTF-based gradient pre-emphasis on the signal phase and consequently on the artifact level in Cartesian bSSFP sequences with three different line-ordering schemes. For evaluation purposes, phantom images were acquired, and the signal-, noise- and artifact levels were assessed quantitatively. The results show that the GSTF based pre-emphasis approach can successfully reduce residual unbalanced gradient artifacts in bSSFP imaging.

Our simulations revealed that for bSSFP imaging with linear ordering of the phase-encoding lines, the additionally acquired phases from eddy currents only change very little between consecutive RF pulses, so the steady-state of the magnetization is almost undisturbed. This result was confirmed by the virtually artifact-free images in Figure 5a. Consequently, the relative signal difference to the linearly ordered image with the GSTF-based pre-emphasis could serve as a quantitative measure of the artifact level. For the centric out and the Niederreiter line-ordering schemes, the simulations exposed the 1$^{st}$ order phase contribution caused by the phase-encoding gradients as the principal source of steady-state disruptions, as its changes from pulse to pulse exceeded all the other phase terms' changes. The artifacts extending in the phase-encoding direction in the phantom images in Figure 5b and c support the existent steady-state disturbances caused by the phase-encoding gradients, also confirming the results of previous studies [14] [15]. Our GSTF-based pre-emphasis greatly reduced these artifacts. Quantitatively speaking, the artifact levels were lowered by 64 % and 89% for centric out and Niederreiter line ordering, respectively. We thus proved that a GSTF-based gradient pre-emphasis can enable artifact-free bSSFP imaging with arbitrary orders of the phase-encoding steps.

In recent studies [9] [10], a GSTF-based gradient pre-emphasis was applied to spiral readout trajectories in cardiac real-time sequences. In the used spoiled gradient echo sequences the GSTF-based pre-emphasis was only applied during the read-out to avoid trajectory errors. The application of a GSTF-based pre-emphasis has also previously been demonstrated to heal geometric distortions in echo-planar read-outs [4]. In principle, all these trajectory deviations can also be dealt with at reconstruction time [5]. In contrast, bSSFP sequences present a more compelling use case for optimized gradient progressions, because non-compensated eddy currents cause disruptions of the steady-state, and cannot be corrected retrospectively.

Several studies have shown that eddy current-induced artifacts in bSSFP imaging can be reduced by a careful choice of the sampling scheme, for example by line pairing [14, 15] or using smoothly interleaved spirals [21]. The premise behind these strategies is to avoid large jumps in k-space. However, this limits the options regarding k-space trajectory design and can possibly limit the scan efficiency. Removing the confounding eddy current and vibration effects by means of an advanced gradient pre-emphasis, can enable artifact-free bSSFP imaging regardless of large jumps in k-space, as our images with the centric out and Niederreiter line-ordering demonstrate.

Such non-linear orders of the phase-encoding lines can be beneficial for accelerated [36, 37] or self-gated imaging [38]. Our results suggest that a GSTF-based gradient pre-emphasis could be an important element to make bSSFP sequences with non-linear line-ordering clinically adaptable.

In GSTF-based simulations, we analyzed the additionally accrued phase due to eddy current- and vibration-induced gradient- and field distortions in our sequence. As the $1^{st}$ order phase contributions were more than ten times larger than the $0^{th}$ order contributions at 10 cm off-center in the non-linear line-ordering schemes, we focused on compensating spatially linear varying fields in our imaging experiments. The applied GSTF-based gradient pre-emphasis not only prevents disturbances of the steady-state, which is achieved by minimizing $d(\Delta\phi)/dn$, but also minimizes the first order contributions to the phase error $\Delta\phi$ itself. Besides establishing a steady-state, this also assures that the same steady-state is reached along a given gradient direction, and position-dependent shifts of the banding artifacts can be avoided.

The main limitations of this study are that we did not test the dependency of the steady-state disruptions on the sequence timing, especially TR, that we did not include a correction for $B_0$ fluctuations, and that we did not validate the results in vivo. Regarding the timing dependency, it is worth noting that the $0^{th}$ and $1^{st}$ order terms of the GSTF have their most prominent resonances at different frequencies, and they exhibit different widths (c.f. Figure 2). Consequently, the oscillating field perturbations have different frequencies and different life times. Therefore, the contributions of these oscillating terms to the phase accrual over each TR critically depends on TR itself. For the sequence timing used specifically in this study, the 1$^{st}$ order phase errors dominated the 0$^{th}$ order phase errors, but this is by no means a general principle. In fact, Bruijnen et al. [15] examined steady-state disruptions in bSSFP sequences that were dominated by the 0$^{th}$ order terms, and proposed RF phase cycling to compensate for $B_0$ fluctuations. Furthermore, the peculiarity of the resonances also depends on the gradient axis and scanner model, so a wider experimental validation of both correction methods should be conducted in the future.

In this work, we did not test correcting eddy current-induced fluctuations of $B_0$. However, correcting $B_0$ eddy currents in addition to gradient fluctuations will almost certainly be necessary to avoid artifacts from steady-state disruptions independent from the sequence timing, scanner model, and image orientation. $B_0$ eddy current corrections for bSSFP imaging have previously been proposed based on adjustments of the phase of the radio-frequency pulses [15, 25]. However, as for the gradient-based, spatially linear contributions, the spatially constant $B_0$–based phase contributions could also be compensated by a continuous, dynamic eddy current correction. This could, for example, be implemented by superimposing a spatially homogeneous, but temporally varying additional field or, more conveniently, by a continuous modulation of the scanner's reference phase [39]. The required knowledge of the dynamic $B_0$ shifts can either be gained through field camera measurements [25] or based on the 0$^{th}$ order terms of the GSTF [15]. We would like to see this point addressed in future work.

Since our advanced gradient pre-emphasis is based on the GSTF, it relies on the assumption that the whole gradient chain is linear and time-invariant. Deviations from this assumption have been reported for temperature changes in the gradient coils [32, 40, 41] and non-linear characteristics of the gradient amplifiers [42]. For the results presented in this work, the measurements of the GSTF and the bSSFP images were acquired on different days (13 months apart) but the GSTF-based pre-emphasis could nevertheless reduce the artifact level in the bSSFP images drastically. However, the impact of deviations of the gradient system from the LTI assumption on bSSFP imaging with a GSTF-based gradient pre-emphasis has yet to be investigated. If the GSTF is also employed to compensate dynamic $B_0$ shifts, the LTI compliance of the respective correction method should also be verified.

Future developments should concentrate on a combination of a prospective gradient pre-emphasis and a continuous, dynamic $B_0$ eddy current correction, both based on the GSTF and directly integrated into the scanner. If the existing, integrated corrections, that model eddy currents only by real exponential functions, could be expanded to rely on complex exponentials and thus integrate oscillatory field fluctuations, scan specific corrections for EPI, or sequences with non-Cartesian trajectories would become obsolete. GSTF-based prospective compensation approaches would allow for improved image quality in many research fields, like compressed sensing [37], or retrospective gating [38]. Furthermore, they would facilitate the translation of topics like non-linear line-ordering in bSSFP imaging, or non-Cartesian trajectories into clinical routine.

## 6 Conclusion

We demonstrated that an advanced gradient pre-emphasis based on the gradient system transfer function can effectively suppress artifacts arising from steady-state disruptions in bSSFP imaging. This opens up the possibilities to employ non-linear Cartesian acquisition schemes in conjunction with undersampling to profit from state-of-the-art acceleration strategies in clinical settings.

**7 Conflicts of interest**

Both the University Hospital Würzburg and the University Clinic Halle/Saale have a research collaboration agreement with Siemens Healthcare GmbH.

**8 Funding**

This study was funded by the German Research Foundation (DFG, grant number: SL 312/1-1), which had no role in the design of the study, in the collection, analyses, or interpretation of data, in writing the manuscript, nor in the decision to publish the results.